# The Near-IR Spectrum of Titan Modeled with an Improved Methane Line List

Jeremy Bailey[a,*], Linda Ahlsved[b], V.S. Meadows[c]

[a]School of Physics, University of New South Wales, NSW 2052, Australia
*Corresponding Author E-mail address: j.bailey@unsw.edu.au
[b]Department of Earth and Planetary Sciences, Macquarie University, NSW 2109, Australia
[c]Department of Astronomy, University of Washington, Box 351580, Seattle, Washington, 9819

ABSTRACT

We have obtained spatially resolved spectra of Titan in the near-infrared J, H and K bands at a resolving power of ~5000 using the near-infrared integral field spectrometer (NIFS) on the Gemini North 8m telescope. Using recent data from the Cassini/Huygens mission on the atmospheric composition and surface and aerosol properties, we develop a multiple-scattering radiative transfer model for the Titan atmosphere. The Titan spectrum at these wavelengths is dominated by absorption due to methane with a series of strong absorption band systems separated by window regions where the surface of Titan can be seen. We use a line-by-line approach to derive the methane absorption coefficients. The methane spectrum is only accurately represented in standard line lists down to ~2.1 μm. However, by making use of recent laboratory data and modeling of the methane spectrum we are able to construct a new line list that can be used down to 1.3 μm. The new line list allows us to generate spectra that are a good match to the observations at all wavelengths longer than 1.3 μm and allow us to model regions, such as the 1.55 μm window that could not be studied usefully with previous line lists such as HITRAN 2008. We point out the importance of the far-wing line shape of strong methane lines in determining the shape of the methane windows. Line shapes with Lorentzian, and sub-Lorentzian regions are needed to match the shape of the windows, but different shape parameters are needed for the 1.55 μm and 2 μm windows. After the methane lines are modelled our observations are sensitive to additional absorptions, and we use the data in the 1.55 μm region to determine a D/H ratio of $1.77 \pm 0.20 \times 10^{-4}$, and a CO mixing ratio of $50 \pm 11$ ppmv. In the 2 μm window we detect absorption features that can be identified with the $\nu_5+3\nu_6$ and $2\nu_3+2\nu_6$ bands of $CH_3D$.

# 1   Introduction

Saturn's satellite Titan has a dense atmosphere composed mostly of nitrogen, but containing methane at an abundance of a few percent. Absorption due to methane is the dominant feature of the near-IR spectrum, giving rise to a series of deep absorption features separated by "windows" where methane is relatively transparent and the surface can be seen. The high methane abundance also results in a methane "hydrological cycle", with the methane condensing as clouds (Griffith et al., 1998; Brown et al., 2002) and probably falling as rain (Tokano et al., 2006) and forming fluvial erosion features on the surface (Jaumann et al., 2008).

Methane is an important atmospheric constituent in many other objects, including the four solar-system giant planets and the brown dwarfs, the presence of methane in the spectrum being the defining characteristic of a T-dwarf. Equilibrium chemistry models (e.g. Lodders and Fegley, 2002) predict that methane will be the main carbon containing species in a solar composition gas mixture at temperature lower than ~1500 K. We therefore expect methane to be present in many of the ~500 extrasolar planets (mostly giant planets) so far detected. Methane has been reported in the spectra of two transiting extrasolar planets (Swain et al., 2008; 2009). For all these objects good models of the spectrum are only possible if accurate spectral line parameters for methane are available.

However because of the complexity of the methane spectrum, standard line lists such as the HITRAN spectroscopic database (Rothman et al., 2005; 2009) only include a good set of methane line parameters for wavelengths longer than about 2.1 µm. At shorter wavelength the methane line data is mostly in the form of room temperature measured parameters that are not useable for a low temperature object such as Titan. Methane absorption data for shorter wavelengths is available in other forms, such as tabulations of absorption coefficients (Karkoschka and Tomasko 2010), band models (Strong et al., 1993) and k-distribution tables (Irwin et al., 2006). However, these are tabulated at 5 – 10 cm$^{-1}$ intervals, and are therefore only suitable for predicting very low resolution spectra. Their application to Titan also involves extrapolation beyond the conditions of the original absorption measurements.

This means, in particular, that the window regions at around 2.0 µm, 1.55 µm and 1.25 µm, between the strong methane bands, cannot be properly studied with high-resolution spectroscopy. However, these are the wavelength regions where we are able to study the surface of Titan, and also the regions where we have the longest optical path through the Titan atmosphere, and hence maximum sensitivity to trace species and lines of methane isotopologues. All such studies are currently severely limited by the inability to fully characterise the methane absorption, and to distinguish other spectral features from the forest of unidentified methane lines. Similar considerations apply to the same wavelength regions in the spectra of other methane atmosphere such as those of the giant planets.

Recently there have been substantial improvements in the available methane line data for the 1.3 to 1.8 µm region, in particular through the application of low-temperature Direct Absorption Spectroscopy (DAS) and Cavity Ring Down Spectroscopy (CRDS) techniques (Campargue et al., 2010a; Wang et al., 2010a, 2010b). Titan is an ideal object to test the new line data, because of its large methane path lengths, and the availability of in-situ measurements from the Huygens probe to constrain the atmospheric structure and properties.

In this paper we use this new low temperature data as well as recent room temperature measurements and identifications (Campargue et al., 2010b, Nikitin et al., 2010a, 2010b) and model predictions (Wenger and Champion, 1998) to construct an improved low-temperature methane line list over the wavelength range from 1.3 to 2.1 µm. We use the new list in conjunction with a line-by-line, multiple scattering, radiative transfer model to predict the Titan spectrum and compare with observed spectra of Titan at a resolving power of ~5000.

We also investigate the effects of the far-wing line shape on the spectrum of Titan in the methane windows. The absorption in these regions has significant contributions from the far wings of strong methane lines some distance away. This effect is known to be important in other similar cases. For example, the $CO_2$ far wing line shape is important in modeling the window regions in the Venus nightside spectrum (Pollack et al., 1993; Meadows & Crisp, 1996) but has not been much investigated in the case of Titan.

We also use our models to investigate the spectra of other features in the 1.55 µm and 2.0 µm windows that can be seen more clearly after the $CH_4$ lines are modeled. We measure the mixing ratio of CO and investigate lines of the methane isotopologue $CH_3D$ in both windows.

## 2  Observations

Titan was observed on 2006 February 2, February 7 and February 8 as part of the System Verification phase of the Near-Infrared Integral Field Spectrometer (NIFS) on the Gemini North 8m telescope at Mauna Kea, Hawaii. NIFS (McGregor et al., 2003) is an instrument built for the Gemini telescope by the Research School of Astronomy and Astrophysics at the Australian National University. NIFS uses an optical system that reformats a 3.0 × 3.0 arc second field into 29 slices, which are then dispersed by a spectrograph, and the resulting spectra are imaged simultaneously onto a Rockwell HAWAII-IIRG 2048 × 2048 HgCdTe detector array. The instrument thus records a 3D spectral cube covering one of the J (1.15 – 1.36 µm), H (1.48 – 1.8 µm) or K (2.01 – 2.43 µm) bands with a single observation. The spectral resolving power is R=5290 in the H and K bands and 6040 in the J band. The K band data from this observation set have also been used for a different investigation by Kim et al. (2008).

NIFS is used in conjunction with the ALTAIR adaptive optics system (Richardson et al., 1998, Herriot et al., 1998), which provides diffraction limited images with a FWHM of better than 0.1 arc seconds. The observations used for this paper were those from 2006 Feb 2, on which all three near-IR bands were observed. Details of the observations are given in Table 1. At the time of observation Saturn was close to opposition and passing almost overhead from Mauna Kea. The phase angle was 0.65 degrees. Titan had an angular diameter of 0.874 arc seconds. The seeing recorded by the ALTAIR system was excellent (0.25 – 0.4 arc seconds). Under such conditions ALTAIR should deliver images with FWHM of ~0.07 arc seconds, and Strehl ratios of ~0.4 at K and ~0.25 at H (Christou et al., 2010). However, the optics of NIFS limit the actual resolution to 0.1 – 0.15 arc seconds.

*Table 1 — Titan Observations on 2006 Feb 2*

| UT (mid Point) | Wavelength | Air-mass | Sub Observer | |
|---|---|---|---|---|
| | | | Longitude | Latitude |
| 10:22 | J | 1.00 | 172.6 | –18.75 |
| 11:15 | H | 1.04 | 173.4 | –18.75 |
| 11:50 | K | 1.09 | 173.6 | –18.75 |

Each observation consisted of a series containing four on-source and four off-source exposures. The exposure times were 180 seconds at J and H and 240 seconds at K. In addition observations were made of standard stars of spectral type G2V. The standard stars were HIP 36874 (V = 7.55, J = 6.168) for J, and HIP 49942 (V = 8.43, H = 6.859, K = 6.808) for H and K. The data were reduced using the Gemini IRAF reduction package for the instrument to generate wavelength calibrated spectral cubes.

Fig. 1 shows image slices from the K band cube at three different wavelengths. The 2.03 μm image is in a methane transparency window and shows surface markings. The 2.12 μm image shows the troposphere. At this wavelength clouds appear as bright features (Roe et al., 2005). The lack of any such features in this image shows that Titan was cloud free at this time at the spatial resolution of our observations. This image and the 2.30 μm image that is sensitive to the stratosphere, show limb brightening due to the aerosol haze.

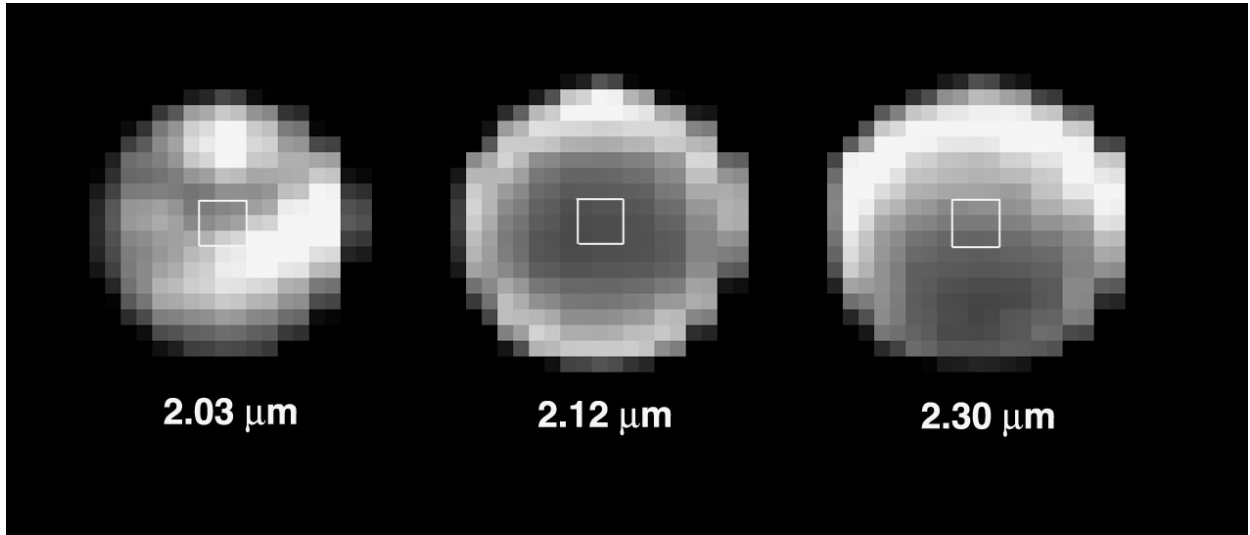

*Figure 1 – Images extracted from the K band spectral cube at three wavelengths. The 2.03 μm image is in a methane transparency window and shows surface markings. The 2.12 μm image is sensitive to the troposphere, and the 2.30 μm image shows the stratosphere. If clouds were present they would show up as bright features in the 2.12 μm image. The absence of such features shows that Titan was cloud free at the time of these observations. The 3 by 3 pixel region extracted to give our spectra is indicated.*

We have extracted from each cube the spectrum summed over the central 3 × 3 spatial pixels where the pixels are 0.05 arc seconds square. It should be noted that because the correction of the adaptive optics is not perfect, the spatial point spread function has an uncorrected halo with the size of the original seeing disk. Thus although much of the light will come from the centre of Titan there will also be a significant contribution from regions of the disk away from the centre These spectra have been divided by the corresponding standard star spectra. Since the standard stars are of solar spectral type (G2V) the division by the standard star largely removes both telluric absorption features and solar spectral lines, and results in a spectrum proportional to the radiance factor (I/F). The NIFS data reduction system does not provide flux calibrated data cubes, so we have adjusted the absolute calibration of the radiance factor to match typical values obtained in past observations of the cloud-free Titan spectrum (e.g. Adamkovics et al., 2010; Schaller et al., 2009).

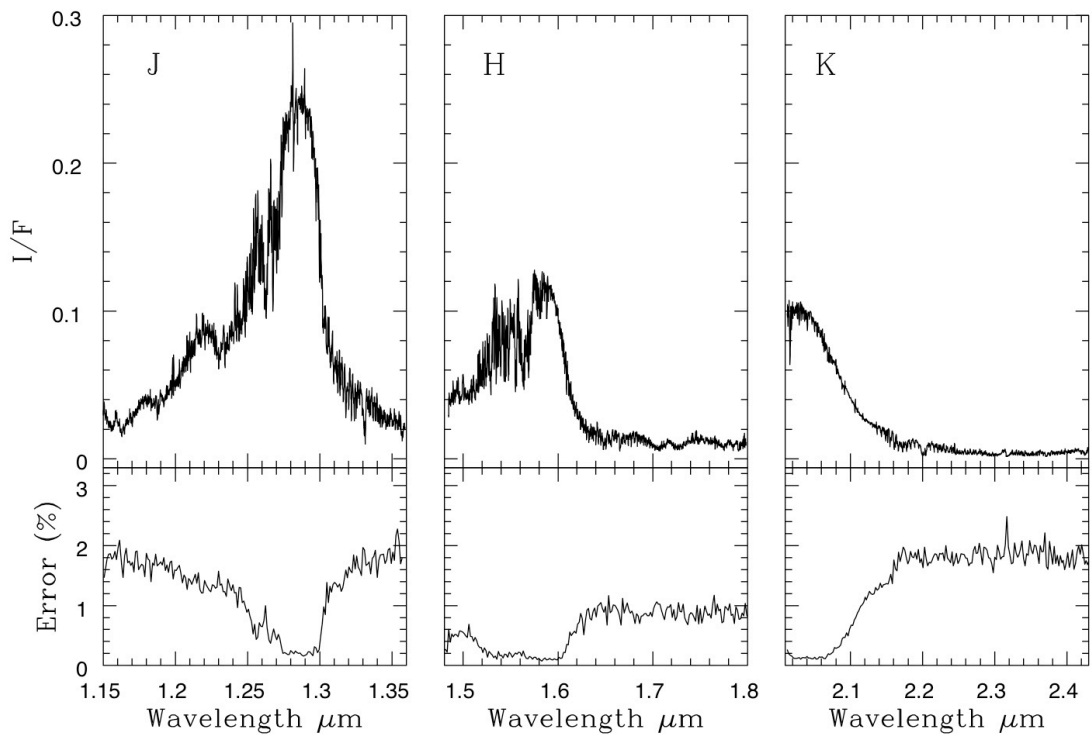

*Figure 2 – NIFS spectra extracted from the centre of the Titan disk for the J, H and K bands. The lower panels show the error on individual spectral pixels determined from the statistics of the four individual observations making up each cube.*

In practice the cancellation of solar features is not perfect because of the different Doppler shifts of the star and the solar lines reflected from Titan. Residual features are apparent on strong solar lines, most obviously the Paschen-β line of hydrogen at 1.28 μm which shows up as a positive spike on the blue side, and a negative spike on the red side. Similar, but weaker features are apparent in the K-band spectrum near 2.1 μm.

The resulting spectra are shown in Fig. 2. These spectra form the basis of the subsequent analysis. The spectra have high signal to noise ratio as shown in the lower panels of figure 2, where we show the percentage error per individual spatial pixel, as derived from a smoothed representation of the statistical variation between the four individual object/sky pairs that make up each data cube. The percentage error ranges from 0.1–0.2 % in the window regions to ~2 % in the deep absorption bands.

## 3  Radiative Transfer Model

We modeled the atmosphere of Titan using the Versatile Software for Transfer of Atmospheric Radiation (VSTAR) software package (Bailey 2006). VSTAR has previously been used for the atmospheres of Earth, Venus and Jupiter (Hough et al., 2006; Bailey et al., 2007, 2008; Bailey, 2009; Kedziora-Chudczer and Bailey, 2011).

The Titan atmosphere was represented with 30 vertical layers from 0 to 200 km, with step sizes ranging from 0.5 km in the lower atmosphere to 15 km in the upper layers. The vertical profile of pressure and temperature was taken from Voyager radio occultation data (Lindal et al. 1983) and is in excellent agreement with in-situ measurements from Huygens (Fulchigoni et al., 2005). The methane volume mixing ratio as a function of altitude is that derived from the Huygens Gas Chromatograph Mass Spectrometer (GCMS, Niemann et al., 2005) and is about 1.5 % for most of the atmosphere increasing to 5 % near the surface.

In addition to methane (discussed in the next section) the only other gaseous absorption components included were carbon monoxide (CO), using line data from HITRAN 2008 (Rothman et al., 2009), and the collision-induced absorption due to an $H_2$ - $N_2$ mixture for which the absorption cross-section was taken from McKellar et al. (1989). The latter is present over the range 4000 — 4900 cm$^{-1}$.

The gaseous absorption, Rayleigh scattering due to $N_2$, and aerosol scattering data (see section 5) are combined to derive the vertical optical depth, single scattering albedo and phase function for each layer of the atmosphere at each spectral bin. The radiative transfer equation is then solved for each spectral bin using an 8 stream discrete ordinate method implemented using the DISORT software package (Stamnes et al., 1988). The up-welling radiance at the top of the atmosphere can then be used to derive a modeled I/F spectrum that can be compared with the observations.

## 4  The Methane Spectrum

A tetrahedral molecule like methane has four vibrational modes: two bending modes described by quantum numbers $\nu_1$ and $\nu_3$ and two stretching modes $\nu_2$ and $\nu_4$, so that each vibrational level can be described by some combination of the four quantum numbers $\nu_1$, $\nu_2$, $\nu_3$, $\nu_4$. For methane the fundamental frequencies of these modes obey the approximate relationship

$$\nu_1 \approx 2\nu_2 \approx \nu_3 \approx 2\nu_4 \approx 3000 \text{ cm}^{-1}$$

These coincidences result in a complex series of interacting vibrational states repeating at intervals of ~1500 cm$^{-1}$ known as polyads. Polyads are generally designated by $P_n$ where n is the polyad number defined in terms of the vibrational quantum numbers by

$$n = 2(\nu_1 + \nu_3) + \nu_2 + \nu_4.$$

$P_0$ denotes the ground state. The first polyad $P_1$ (or the Dyad) contains the two vibrational levels $v_2$ and $v_4$. The $P_2$ polyad (or Pentad) contains five vibrational levels $v_1$, $2v_2$, $v_3$, $2v_4$ and $v_2+v_4$. Successively higher polyads contain rapidly growing numbers of levels. In addition degeneracies of some of the modes lead to a number of sublevels. The polyad structure is described further in Table 2.

*Table 2 — The vibrational polyads of methane in the near infrared*

| Polyad | Name | Approx energy (cm$^{-1}$) | Vibrational Levels | Sublevels |
|---|---|---|---|---|
| $P_7$ | Tetracontad | ~10200 | 40 | 538 |
| $P_6$ | Triacontad | ~8800 | 30 | 280 |
| $P_5$ | Icosad | ~7300 | 20 | 134 |
| $P_4$ | Tetradecad | ~5900 | 14 | 60 |
| $P_3$ | Octad | ~4400 | 8 | 24 |
| $P_2$ | Pentad | ~3000 | 5 | 9 |
| $P_1$ | Dyad | ~1500 | 2 | 2 |
| $P_0$ | Ground State | 0 | 1 | 1 |

Transitions between the higher polyads and the ground state lead to complex overlapping band systems, and interactions between levels distort line positions making these bands very complex to analyse, particularly at higher frequencies. Transitions between higher levels (hot bands) overlay these bands and further complicate the spectrum at higher temperatures (for example the Pentad – Dyad system occurs at the same wavelengths as the Dyad – GS system).

## 4.1 Methane Line Data in the HITRAN Database

A standard spectral line database for atmospheric modeling is the HITRAN database. The latest edition, HITRAN 2008 (Rothman et al., 2009) includes a good set of methane line data for the levels up to the octad, covering frequencies from 0 to 4800 cm$^{-1}$. This part of the methane spectrum is well modeled. However, the line data included in HITRAN for frequencies above 4800 cm$^{-1}$ are empirical line parameters, mostly from Brown (2005). These are room temperature measurements of line position and intensity, but mostly lack quantum identifications and lower state energies.

The lack of a lower state energy is a major problem for modeling a low temperature object like Titan. Line intensities in HITRAN are listed for a reference temperature of $T_{ref}$ = 296 K. The line intensity $S(T)$ at a different temperature is then obtained using (Rothman et al., 1998; Appendix A):

$$S(T) = S(T_{ref}) \frac{Q(T_{ref})\exp(-c_2 E_l/T)(1-\exp(-c_2 v/T))}{Q(T)\exp(-c_2 E_l/T_{ref})(1-\exp(-c_2 v/T_{ref}))} \qquad 1$$

where $Q(T)$ is the partition function (this can be obtained from models provided with the HITRAN database due to Fischer, et al. 2003), $E_l$ is the lower state energy in cm$^{-1}$, $v$ is the line centre frequency in cm$^{-1}$ and $c_2$ ($= hc/k$) is the second radiation constant. The Boltzmann factor terms in this expression make the line intensity strongly dependent on the lower state energy at temperatures far from the reference temperature.

For transitions to the ground vibrational state, the lower state energy is determined by the rotational quantum number J. Thus if the quantum levels of the transition are identified the lower state energy is known. However, where empirical line measurements without identifications are used, the line intensity may be well determined at 296 K, but cannot be reliably converted to a low temperature value for modeling an object like Titan. HITRAN 2008 actually lists fictitious lower state energies of 555.5555 or 333.3333 $cm^{-1}$ for these lines, which enables an intensity to be calculated for a different temperature, but the value is not likely to be meaningful if the temperature is far from 296 K.

Another issue with the HITRAN line data for frequencies above 4800 $cm^{-1}$ is that the intensity cut off is too high to include weak lines, which are significant at the high methane path lengths encountered on Titan. This is a particular problem in the transparency windows between the strong methane band systems.

## 4.2 The Spherical Top Data System (STDS)

STDS is a software package developed at the Université de Bourgogne to simulate the spectra of spherical top molecules such as methane (Wenger and Champion 1998). STDS predicts line positions and intensities using a parameterised representation of the effective hamiltonian and effective dipole moment. These parameters are in turn determined by fitting a large set of experimental line positions and intensities. STDS is widely used for analysis of the methane spectrum and indeed many of the lower frequency line parameters in HITRAN are calculated using STDS. The HITRAN 2004 line data (Rothman et al., 2005) were based on a combination of empirical line data and separate STDS analyses of each polyad system (e.g. Hilico et al. 2001 for the octad; Hilico et al. 1994 and Fejard et al. 2000 for the pentad).

Recently a new global analysis of methane line data from 0-4800 $cm^{-1}$ (all levels up to the octad) has been carried out which provides a greatly improved fit of line positions and intensities (Albert et al. 2009). For example line positions in the octad have an RMS of 0.0035 $cm^{-1}$ compared with 0.041 $cm^{-1}$ for the analysis of Hilico et al. (2001). The parameters of this new fit are incorporated in the current release of STDS and were used as the basis for much of the methane line data included in HITRAN 2008 (Rothman et al., 2009).

This model together with empirical line data provides a good description of the methane spectrum for frequencies up to 4800 $cm^{-1}$. STDS can also be used in the Tetradecad region (5000–6800 $cm^{-1}$) but only a few of the bands that make up this complex system have been analysed and the model is a preliminary one (Boudon et al., 2006; Robert et al., 2001). Line positions and intensities predicted by STDS in this region do not always agree well with laboratory data.

## 4.3 Empirical Line Data and Lower State Energies

Where no quantum identifications exist, it is possible to obtain an empirical value for the lower state energy if the line intensity can be measured at two different temperatures. HITRAN 2008 includes some empirical lower state energies for lines in the 5500 – 6150 cm$^{-1}$ region from Margolis (1990), and in the 5852 – 6181 cm$^{-1}$ region from Gao et al. (2009).

Recently several improved sets of empirical line data for methane have become available (Campargue et al., 2010a, 2010b; Wang et al., 2010a, 2010b). These are based on high sensitivity Direct Absorption Spectroscopy (DAS) or Cavity Ring Down Spectroscopy (CRDS) and are significantly deeper than previous measurements, and include low temperature measurements at ~80 K. In addition, a new methane line list in the 5550 – 6236 cm$^{-1}$ region has been established for the GOSAT project (the GOSAT-2009 list, Nikitin et al., 2010a). There have also been some improvements in identification of methane lines in the 6287 – 6550 cm$^{-1}$ region (Nikitin et al., 2010b).

Using this new data we have been able to generate a new methane line list, suitable for modeling the Titan spectrum down to a wavelength of 1.3 μm (7655 cm$^{-1}$).

## 4.4 Construction of the new Methane Line List

The sources of line data used in our new line list are summarized in Table 3 and described in more detail below.

Methane line data in the region 0 – 4800 cm$^{-1}$ were taken from HITRAN 2008 (Rothman et al., 2009). Since a good model for the methane spectrum is available in this region, most line parameters in this region are either based on the global model of Albert et al. (2009), or on high-quality empirical data with good quantum identifications. Lines of isotopologues of methane in this region are also included.

*Table 3 – The Methane Line List*

| Wavenumber Range (cm$^{-1}$) | Wavelength Range (μm) | Source of line data | Comments |
|---|---|---|---|
| 0 – 4800 | > 2.0833 | HITRAN 2008 (1) | Well modeled region |
| 4700 – 5500 | 1.818 – 2.128 | STDS (2) | Preliminary model for Tetradecad |
| 5500 – 5550 | 1.802 – 1.818 | HITRAN 2008 (1) | HITRAN includes empirical lower state energies from ref 3 |
| 5550 – 6236 | 1.604 – 1.802 | GOSAT-2009 (4) | Supplemented with low-temperature line data from ref 5. |
| 6236 – 6289 | 1.590 – 1.604 | CRDS (6) | Some line identifications from ref 7 |
| 6289 – 6526 | 1.532 – 1.590 | CRDS (8) | |
| 6526 – 6717 | 1.489 – 1.532 | CRDS (6) | Some line identifications from ref 7 |
| 6717 – 7655 | 1.300 – 1.489 | DAS (9) | |
| 7655 – 9200 | 1.087 – 1.300 | HITRAN 2008 (1) | Based on data from ref 10 |

References to Table 3

1. Rothman et al. (2009)
2. Wenger & Champion (1998)

3   Margolis (1990)
4   Nikitin et al. (2010a)
5   Wang et al. (2010a)
6   Campargue et al. (2010b)
7   Nikitin et al. (2010b)
8   Wang et al. (2010b)
9   Campargue et al. (2010a)
10  Brown (2005)

Line parameters from 4700 – 5500 cm$^{-1}$ were calculated using STDS (Wenger and Champion 1998) based on the model parameters for the Tetradecad provided with that software. While this model is a preliminary one based on measurements of very few of the bands that make up the Tetradecad, it is preferred to the empirical data included in HITRAN 2008 in this region, because all its lines have lower state energies, and hence can be reliably converted to Titan temperatures. In this region many of the strongest lines are in the $4\nu_4$ band system for which measurements and analysis have been presented by Robert et al. (2001). There is an overlap between line data from the Tetradecad model and from HITRAN in the 4700–4800 cm$^{-1}$ range, but the HITRAN data are all for the $P_3$ polyad, while the model data are all for the $P_4$ polyad, so there is no duplication. Lines down to an intensity of $10^{-28}$ cm mol$^{-1}$ at 296 K were included in the list. Line parameters from 5500–5550 cm$^{-1}$ were taken from HITRAN 2008. In this region empirical lower state energies for most of the lines from Margolis (1990) are included in HITRAN.

The basis of line data in the range 5550 – 6236 cm$^{-1}$ is the GOSAT-2009 list (Nikitin et al., 2010a). To this we have added 306 empirical lower state energies from Wang et al. (2010a), which were based on a comparison of the GOSAT-2009 list with low-temperature (81 K) measurements, and we have added a further 889 lines measured at low temperature by Wang et al. (2010a) which have no counterparts at room temperature. For these lines we assumed that the failure to detect the line at room temperature was due to a low value for the lower state energy, and therefore assigned a lower state energy value of 100 cm$^{-1}$. Because these lines are measured at 81 K, a temperature close to that of the Titan atmosphere, the error arising from uncertainty in the lower state energy is much smaller than would be the case for a line measured at room temperature.

Line parameters from 6289 – 6526 cm$^{-1}$ (the transparency window between the Tetradecad and Icosad) were based on low temperature (80 K) CRDS measurements by Wang et al. (2010b). We used empirical lower state energies where available, and where lines were detected only in the low temperature measurements we assumed a lower state energy of 100 cm$^{-1}$, as in the Tetradecad region. In this region a significant number of lines are due to CH$_3$D, which makes a substantial contribution to absorption at the transparency window as pointed out by Wang et al. (2010b). The CH$_3$D lines are incorporated in the line list with their intensities scaled by the isotopologue's terrestrial abundance as is the convention in HITRAN and similar line lists. The intensities will need to be scaled if a different D/H ratio is required.

Lines from 6717 – 7655 cm$^{-1}$ (the Icosad) were taken from low temperature measurements by Campargue et al. (2010a) using the same procedures as in the lower frequency regions. This leaves two small regions (6181 – 6289 and 6526 – 6717 cm$^{-1}$) where no low temperature measurements are available. In these regions we used room temperature measurements by Campargue et al. (2010b). Some lines in these regions were assigned lower state energies based on quantum identifications by Nikitin et al. (2010b). However, most strong lines in these two regions have no lower state energy.

Given that these regions are relatively narrow it is possible to make an estimate of the typical value of lower state energy by looking at the general pattern of variation of lower state energy with frequency. Fig. 3 shows the intensity-weighted mean value of lower state energy from our line list in 10 cm$^{-1}$ bins from 4000 to 7655 cm$^{-1}$. Considerable structure is apparent in this diagram. The averaged lower state energy is low in regions of strong methane absorption, and increases in the more transparent regions between the strong band systems, where more outlying high-J transitions are likely to be encountered. On the basis of this diagram we have adopted values of lower state energies of 230 cm$^{-1}$ for the region 6526–6717 cm$^{-1}$ and 510 cm$^{-1}$ for the region 6181 – 6289 cm$^{-1}$. These values are assigned to all lines in these regions that otherwise lack a lower state energy determination. While these values are likely to be representative of the mean value of lower state energy in these regions, the values of individual lines are still expected to vary widely, so we can't expect individual line intensities in these regions to be accurately modeled.

There are no new data available at wavelengths shorter than 1.3 µm but we have included in our list lines from 7655 cm$^{-1}$ to 9200 cm$^{-1}$ taken from HITRAN 2008, which are room temperature measurements from Brown (2005). Apart from a few identifications in the 3ν$_3$ band at around 9000 cm$^{-1}$ the lines lack lower state energies. There are no lines between 7698 and 8002 cm$^{-1}$.

The line data taken from HITRAN 2008 and GOSAT-2009 include values for the air-broadened and self-broadened line width coefficients ($\gamma_{air}$ and $\gamma_{self}$ in cm$^{-1}$ atm$^{-1}$) and an exponent (*n*) for the temperature dependence of the line width. The data from other sources do not include line width data, and for these lines we adopted the mean values from Nikitin et al. (2010a) of $\gamma_{air}$ = 0.060, $\gamma_{self}$ = 0.077 and *n* = 0.85. The broadening gas in the Titan atmosphere is N$_2$ rather than air. From measurements of O$_2$ and N$_2$ broadening of CH$_4$ lines by Lyulin et al. (2009) it is found that $\gamma_{N2}$ is larger than $\gamma_{air}$ by typically 1 to 1.5%. This is, in most cases, much less than the uncertainty in the line width data and we have therefore used the $\gamma_{air}$ values without any correction.

Data on CH$_4$ line broadening in different broadening gases by Pine (1992) and Pine and Gabard (2003) shows that line widths broadened by H$_2$ are about 2% higher than those broadened by N$_2$, while those in He are about 35% lower. This indicates that the line widths in a typical H$_2$/He giant planet atmosphere should be little different to those in air, and our line list should be useable for such atmospheres without any correction.

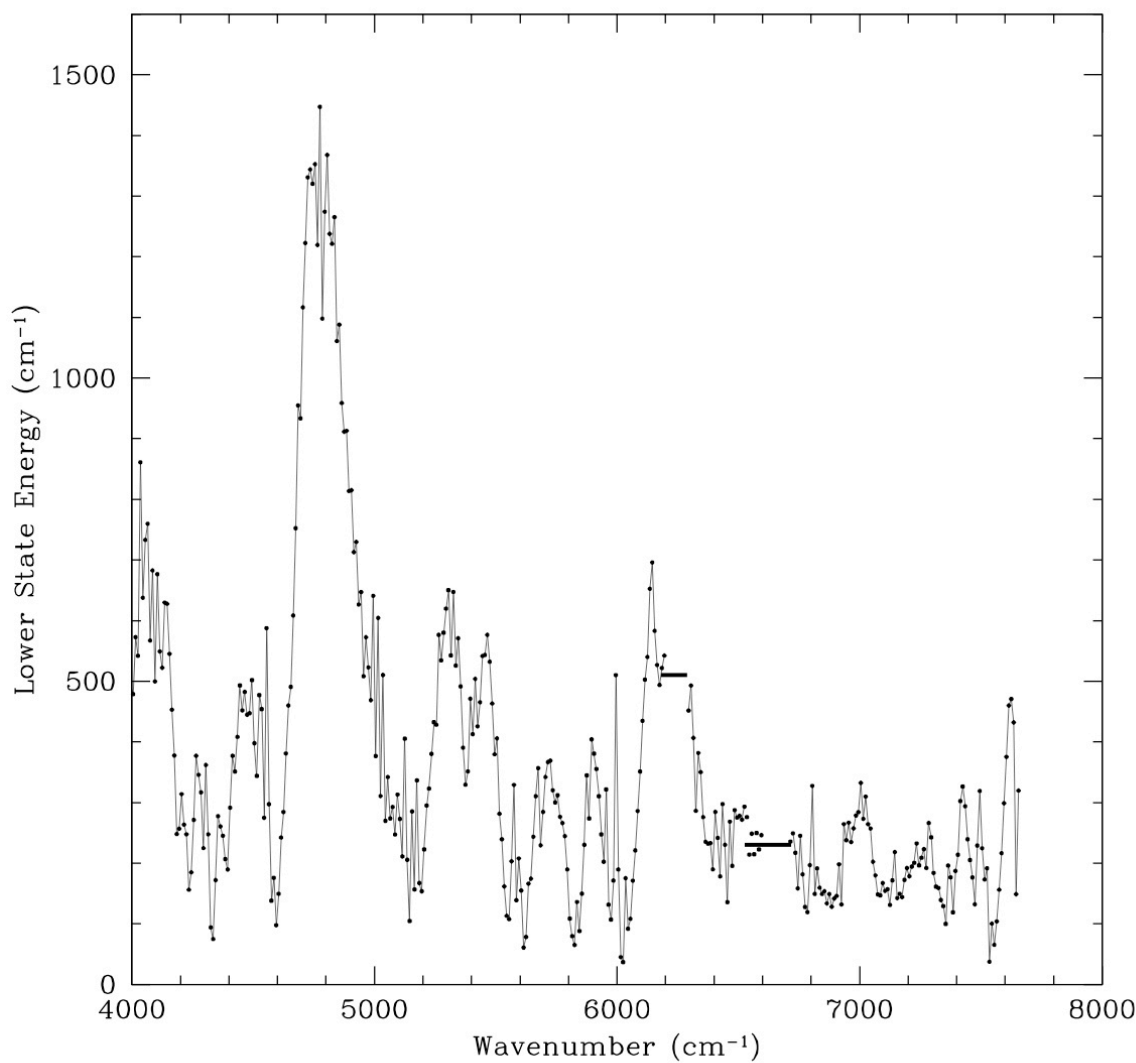

*Figure 3 – Intensity weighted average values of lower state energies for spectral lines in our line list combined in 10 cm$^{-1}$ bins. Horizontal lines indicate our adopted values of lower state energy for lines in the regions 6181 – 6289 cm$^{-1}$ and 6526 – 6717 cm$^{-1}$.*

## 4.5 Comparison with HITRAN 2008

Fig. 4 shows a comparison of the new line list with the methane line data in HITRAN 2008 over the wavelength range 1.1 to 2.0 μm. The 296 K line intensity is plotted against wavelength. On this diagram dots are lines with no lower state energy, open squares are lines with well determined lower state energies based on quantum identifications or empirical values and crosses are lines with estimated lower state energies as described in the previous section. Not only is the new list deeper and more complete, but a much higher fraction of the lines have lower state energy determinations. In HITRAN only 14% of the 42817 lines over 1.1 to 2.0 μm have lower state energies. In the new list 52.8% of the 85538 lines over the same range have lower state energies, and a further 19.4% have estimated lower state energies as described in the previous section. The remaining lines without lower state energies are generally the weaker lines and those at wavelengths below 1.3 μm where no new data is available.

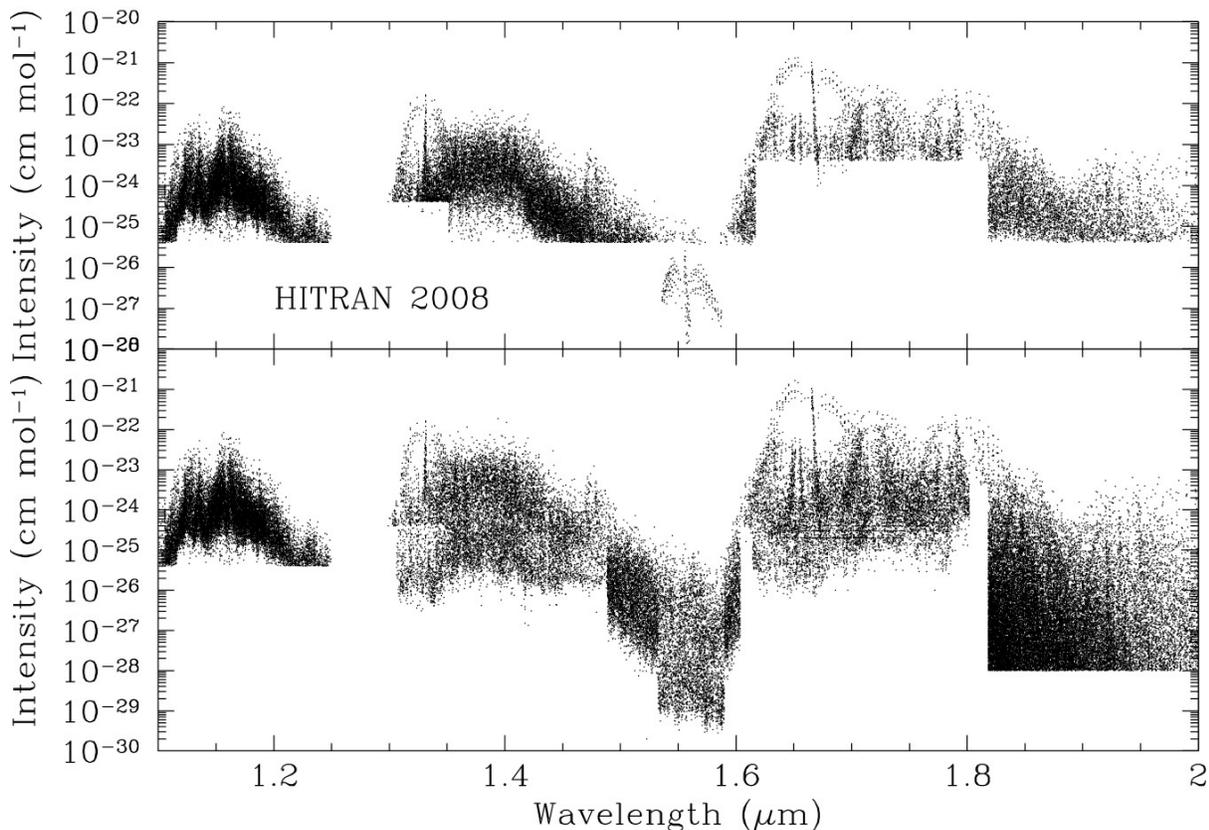

*Figure 4 – Comparison of the new methane line list (lower panel) with the HITRAN 2008 list (upper panel). Line intensity at 296 K is plotted against wavelength.*

A particular advantage of the new list is that it includes many more weak lines in the 1.6 μm transparency region due to the new CRDS data from Campargue et al. (2010b) and Wang et al. (2010b). In this region HITRAN 2008 has an intensity limit of $3.7 \times 10^{-26}$ cm mol$^{-1}$ and this is not deep enough to include a significant number of methane lines, but it is at these wavelengths where we see the most flux from Titan. The lines in HITRAN at around 1.56 μm with intensities from $10^{-26}$ to $10^{-28}$ are due to the $3\nu_2$ band of the isotopologue $CH_3D$. This is actually a strong band, but following the HITRAN convention its line intensities are scaled by the isotopologue's terrestrial abundance of $6.15751 \times 10^{-4}$. The transparency window at 1.25 μm is still not covered by the new line list.

## 5   Aerosol Model and Surface Albedo

### 5.1   Aerosol Model

Titan's atmosphere includes a haze of aerosol particles that obscure the view of the surface at visible wavelengths, but become more transparent in the near-IR. A good model for the scattering by these aerosol particles is essential for radiative transfer modeling of the atmosphere.

Tomasko et al. (2008) have developed a detailed model for the aerosol properties based on measurements made with the descent imager/spectral radiometer (DISR) during the Huygens probe's descent through the atmosphere. We used this model as the basis for our analysis. The model adopts different aerosol properties for three different altitude ranges (0–30 km, 30–80 km and >80 km). We adopt the power law wavelength dependence of the aerosol opacity for each of these regions given in table 3 of Tomasko et al. (2008), and the wavelength dependence of single-scattering albedo given in table 2 of the same paper. The values are given up to 1.6 μm, the longest wavelength observed by DISR. We assumed that the power laws could be extrapolated to the longer wavelengths covered in our study, and that the single-scattering albedos were constant at wavelengths greater than 1.6 μm. Tomasko et al. (2008) fit the scattering phase functions with models based on fractal aggregates of many small monomers and present tabulations of these phase functions at a range of different wavelengths.

In our modeling we used a simpler representation of the phase function as a Henyey-Greenstein function (Henyey & Greenstein, 1941), which is described by a single asymmetry parameter ($g$) that measures the degree of forward scattering. We determined g for each wavelength by fitting the Henyey-Greenstein function to the phase functions tabulated by Tomasko et al. (2008).

The vertical dependence of aerosol optical depth is also given by Tomasko et al. (2008). In our analysis we allowed the aerosol optical depth from this model to be modified by scaling factors for each of the three altitude regions, which we adjusted to match the observations.

## 5.2 Surface Albedo

The surface of Titan is represented in our model as a Lambertian reflector at the base of the atmosphere. The wavelength dependence of the surface albedo is a smoothed version of that determined by the Huygens probe DISR (Tomasko et al., 2005), decreasing from about 0.2 at 0.8 μm to 0.08 at 1.7 μm and assumed constant at longer wavelengths (which were not observed by DISR). To allow for differences between the Huygens landing site and the surface region covered by our observations we allow the surface albedo to be scaled by a constant factor, which is adjusted to fit the observations.

More recent analyses of DISR data have indicated rather higher surface reflectivities (Schroder and Keller, 2008; Keller et al., 2008) but these higher values are hard to reconcile with observations from outside the atmosphere. Keller et al. (2008) also find differences between the surface reflectivity in the lake bed where Huygens landed and in surrounding land areas. The lake bed reflectivity shows a shallow absorption feature near 1.55 μm which might be due to water ice. The surrounding areas do not show this feature. We used a wavelength dependence of surface albedo that did not include any such absorption features.

## 6 Modeling Results

## 6.1 Comparison of Modeled and Observed Spectra

We initially tried to fit our model to the data by adjusting two scaling factors for the aerosol optical depth at all levels and for the surface albedo. Varying the aerosol scaling factor changes the I/F in the methane absorption bands where the light we see is mostly from aerosol scattering.

Varying the surface albedo changes the I/F in the methane windows only, where we are seeing light reflected from the surface. By adjusting these two quantities we could obtain a model that correctly fitted the I/F levels in both window regions and methane bands, but did not correctly match the depth of the absorption features in either of these two regions, producing methane absorption features that were too shallow in the windows, and slightly too deep in the methane bands.

The depth of the absorption features in the windows is a consequence of the balance between radiation reflecting from the surface, which passes through a large methane path length, and radiation scattering from aerosols higher in the atmosphere that passes through a smaller methane path, and thus acts to dilute the strength of methane absorption features. To match the spectra across all wavelengths, we found it was necessary to apply different scaling factors for the aerosol optical depth in the three altitude regions modeled by Tomasko et al. (2008) of below 30 km, 30 to 80 km and above 80 km. By varying these four quantities (albedo scaling factor, and aerosol scaling factor at three levels) we obtained a model that accurately reproduced the I/F levels in both windows and methane absorption bands, and the methane absorption line depths in both windows and methane absorption bands. From the H-band spectra we derived an albedo scaling factor of 1.2, and aerosol scaling factors of 0.3, 0.5 and 0.8 for the 0 to 30 km, 30 to 80 km and above 80 km levels, with an estimated uncertainty of 0.1. The uncertainties are based on eye estimates of the degree of change in these parameters that left obvious systematic residuals of spectral features in the fits. Note that since we do not have absolutely calibrated data, and all these four parameters depend on the absolute flux level, the real uncertainties in these parameters are certainly larger. The values we have adopted yield an internally consistent model that allows us to proceed with our analysis, but reliable absolute measurement of these parameters could only be made from data with a good absolute flux calibration. The resulting model parameters are summarised in Table 4.

The other factors that influence the resulting spectrum are the far-wing line shape described in the next section, and the CO mixing ratio and D/H ratio as described in section 6.3.

Fig. 5 shows the model spectrum compared with the observations with these model parameters. Also shown on Fig. 5 is the model prediction if the HITRAN 2008 line list is used in place of our new line list. It can be see that the new line list results in a model spectrum that is a good overall match to the observations in the region from 1.3 µm to 2.43 µm. In particular the new line list provides a reasonably good match to the observed shape of the 1.55 µm window. The HITRAN 2008 list fails completely to match the structure in the 1.55 µm window, and is also a poorer match to the data at other wavelengths. It shows, for example, lines in the 2 to 2.1 µm region that are not in the data, and is a poorer match to the data in the 1.6 to 1.8 µm region. This can be understood as a consequence of poor prediction of line intensities at low temperature due to the lack of well-determined lower state energies.

Fig. 6 shows an expanded view of the observed spectrum in the H window compared with the model spectrum. This is the region of the spectrum that benefits most from the new data included in our line list. In general the detailed structure of the observed spectrum is well reproduced by the model. However, there are two regions (from about 1.59 – 1.62 µm and 1.49 – 1.53 µm) where low temperature line measurements were not available, and the line list is instead based largely on room temperature measurements and estimated lower state energies. Not surprisingly the detailed match between the observed and model spectrum is not so good in these regions.

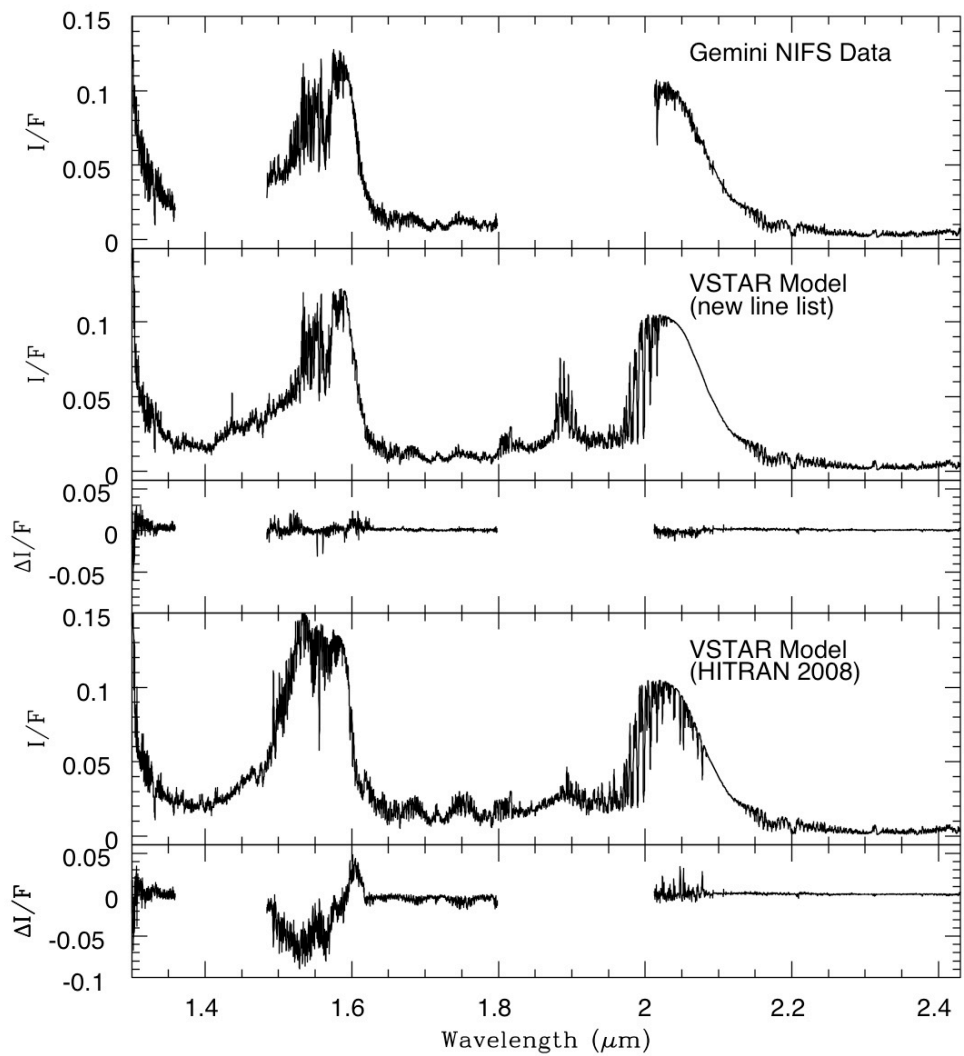

*Figure 5 – Observed spectra of Titan (top panel) compared with predicted spectra using the VSTAR model and the new line list (middle panels), and the same model using the HITRAN 2008 line list (lower panels). The residuals (Data – Model) are shown beneath each model spectrum.*

The lower panel of Fig. 6 shows the residuals of the model fit as a percentage of the modeled spectrum. The RMS residual over the 1.65 to 1.75 micron region (in the methane absorption band) is 7.7%. The residual in the 1.53 to 1.59 micron region in the methane window where we have good low temperature line data is 2 to 2.5 % as discussed further in section 6.3. While this is a substantial improvement on what has been possible in the past, it is still not as good as the measurement errors in these regions, which are about 1% in the absorption band and about 0.2% in the window region. This indicates that the quality of the line list, rather than measurement error is still the limiting factor in our ability to model the observed spectra. It can also be seen that there are places in the spectrum that show residuals of 20% or more, indicating that there are regions of the line list which still need improvement.

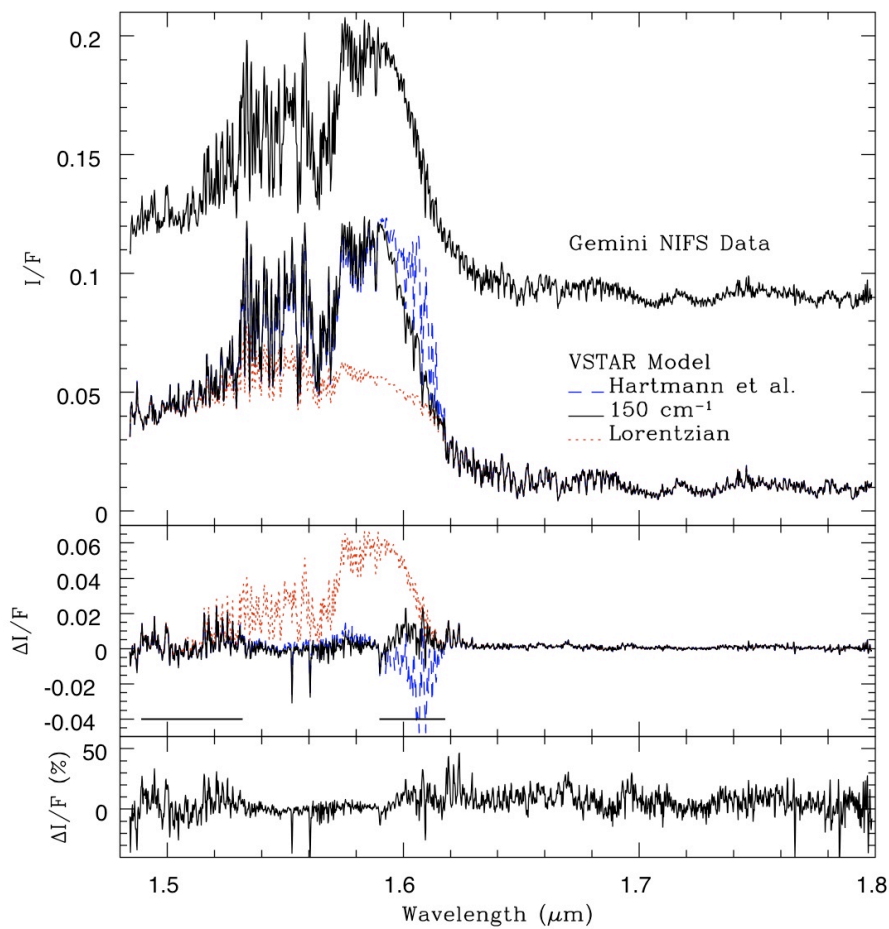

*Figure 6 – Expanded view of the observed spectrum of Titan (shifted up by 0.08) in the H-band (1.48 – 1.8 μm) compared with the model spectrum using the new line list. The middle panel shows the residuals (Data – Model). Horizontal lines show the regions where the line list is less reliable due to the use of estimated lower state energies. The dashed line (blue) uses the line shape model of Hartmann et al. (2002). The dotted line (red) uses a Lorentzian profile continued to large distances from the core. The solid line uses the line shape given in equation 2, which is Lorentzian to 150 $cm^{-1}$, and then becomes sub-Lorentzian. The lower panel shows the residuals as a percentage of the modelled spectrum.*

Fig. 7 shows a similar comparison for the J band (1.15 – 1.36 μm). In our new line list only the region from 1.3 – 1.36 μm has been updated, and it can be seen that this part of the model spectrum matches the observations well. Below 1.3 μm our model fails to match the structure of the 1.25 μm window region due primarily to missing lines in the window region. Fig. 8 shows the comparison for the K band (2.01 – 2.43 μm). The model here uses the same aerosol and albedo scaling factors determined for the H-band, and shows that these factors do a very good job of fitting the K-band spectrum as well. This suggests that our extrapolation of the power law models for the wavelength dependence of aerosol properties as discussed in section 5.1 is a good approximation.

*Table 4 — Model Parameters*

| Parameter | Value | Uncertainty |
|---|---|---|
| Surface Albedo (at 1.6 μm) | 0.108 | 0.009[a] |
| Aerosol optical depth (>80 km) at 1.5 μm | 0.31 | 0.04[a] |
| Aerosol optical depth (30–80 km) at 1.5 μm | 0.34 | 0.07[a] |
| Aerosol optical depth (<30 km) at 1.5 μm | 0.16 | 0.05[a] |
| D/H | $1.77 \times 10^{-4}$ | $0.20 \times 10^{-4}$ |
| CO mixing ratio (ppmv) | 50 | 11 |

a — Uncertainties based on our adopted I/F calibration. The absolute uncertainty in these parameters will be larger.

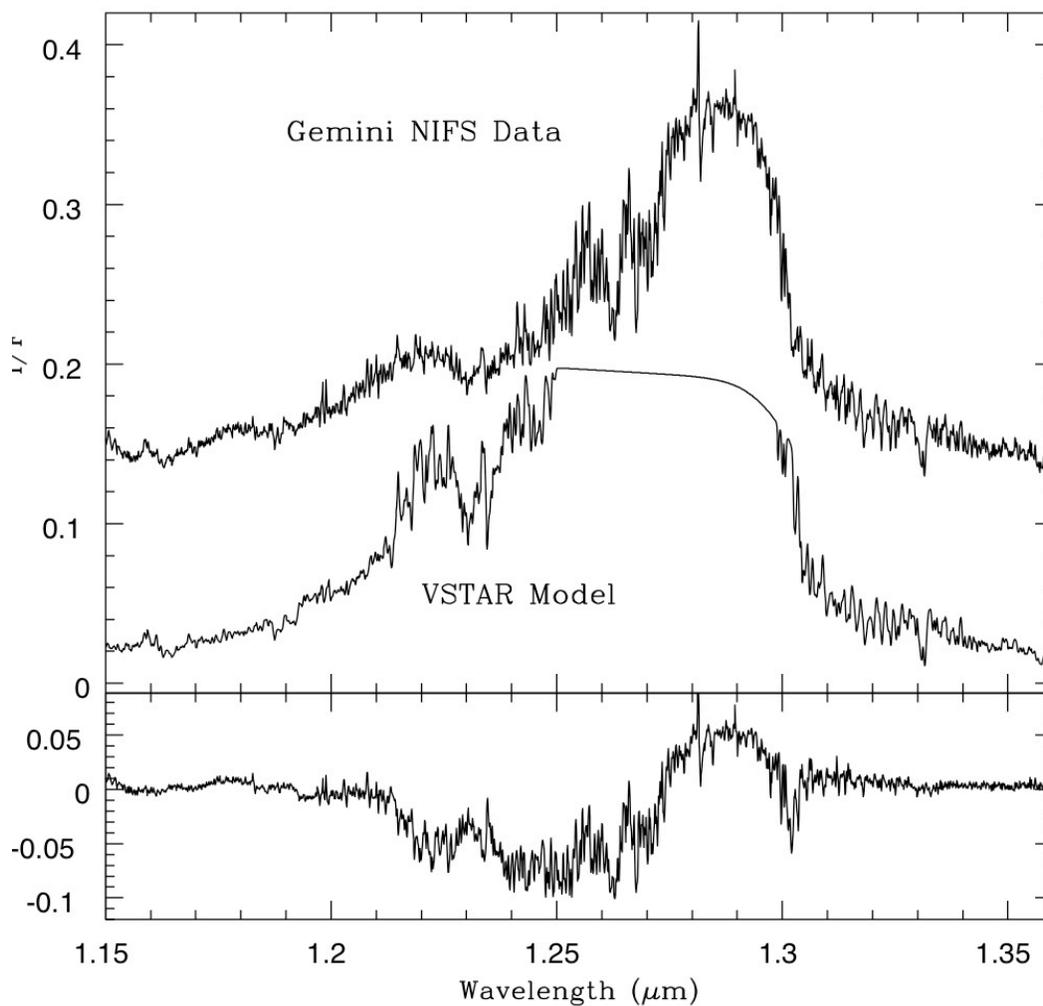

*Figure 7 – The observed spectrum of Titan in the J band (1.15 – 1.36 μm) compared with the model spectrum. Only the region from 1.3 – 1.36 μm has been updated in the new line list.*

## 6.2 Far Wing Line Shape

An important factor in the window regions between the strong methane bands is the shape of the far wings of the methane lines. Where there are few methane lines, the far wings of strong lines some distance away can contribute significantly to the absorption coefficient. The default line shape model in VSTAR uses a Voigt profile in the line core, and a Van Vleck-Weisskopf profile (Van Vleck and Weisskopf, 1945) in the wings. The Van Vleck-Weisskopf profile is a modification of the Lorentzian profile that introduces an asymmetry far from the line centre.

It is often found that the absorption in the far line wings is less than would be predicted by continuing a Lorentzian (or Van Vleck-Weisskopf) profile out to large distances from the line centre. This can be allowed for by modifying the absorption in the line wings by a factor called a $\chi$ factor which is a function of distance from the line centre. Here we refer to a line shape with $\chi = 1$ as a Lorentzian, and $\chi < 1$ as sub-Lorentzian, although strictly in our modeling we use the Van Vleck-Weisskopf rather than the Lorentz profile. The need for such profiles has been investigated, in particular, for $CO_2$, and sub-Lorentzian line shape models have been studied in the laboratory (Perrin & Hartmann, 1989; Tonkov et al. 1996) and found to be necessary to match the shape of the observed windows in Venus near-IR nightside spectrum (Pollack et al., 1993; Meadows & Crisp, 1996). When a $\chi$ factor is applied we also apply a correction to ensure that the integrated line intensity is unchanged.

A sub-Lorentzian line shape model for methane broadened by $H_2$ in the $v_3$ band (in the 3 μm region) has been presented by Hartmann et al. (2002). However, we found this line shape model did not provide a good fit to the shape of the Titan 1.55 μm window. This is probably not surprising in that we are investigating both different bands, and a different broadening gas in the Titan case. A Lorentzian line shape extended to large distances from the line centre however, produces too much absorption in the centre of the window. We found that a reasonable match to the 1.55 μm window was provided by a line shape that was Lorentzian to 150 cm$^{-1}$, and sub-Lorentzian at larger distances as follows:

$$\begin{aligned} 0 < \sigma < 150 cm^{-1} & \quad \chi = 1 \\ 150 < \sigma < 300 cm^{-1} & \quad \chi = \exp(-0.03(\sigma - 150)) \\ \sigma > 300 cm^{-1} & \quad \chi = 0 \end{aligned} \qquad 2$$

where $\sigma$ is the distance from the line centre. This is the line shape used for the models described above. Fig. 6 shows the effects of the different line shapes on the 1.55 μm window.

However, the same line shape does not provide a good match to the shape of the 2 μm window. A Lorentzian line shape produces too much absorption in this window as shown in Fig. 8. The line shape of equation 2, that fits the 1.55 μm window does not fit well. The Hartmann et al. (2002) line shape is even poorer. A better fit to the shape of this window is provided by a line shape that is Lorentzian out to 250 cm$^{-1}$, and is sub-Lorentzian at greater distances as follows:

$$0 < \sigma < 250 cm^{-1} \quad \chi = 1$$
$$250 < \sigma < 400 cm^{-1} \quad \chi = \exp(-0.03(\sigma - 250)) \quad\quad 3$$
$$\sigma > 400 cm^{-1} \quad \chi = 0$$

In the data plotted in Fig. 5 we used the line shape of equation 2 below 2 μm, and the line shape of equation 3 at wavelengths longer than 2 μm.

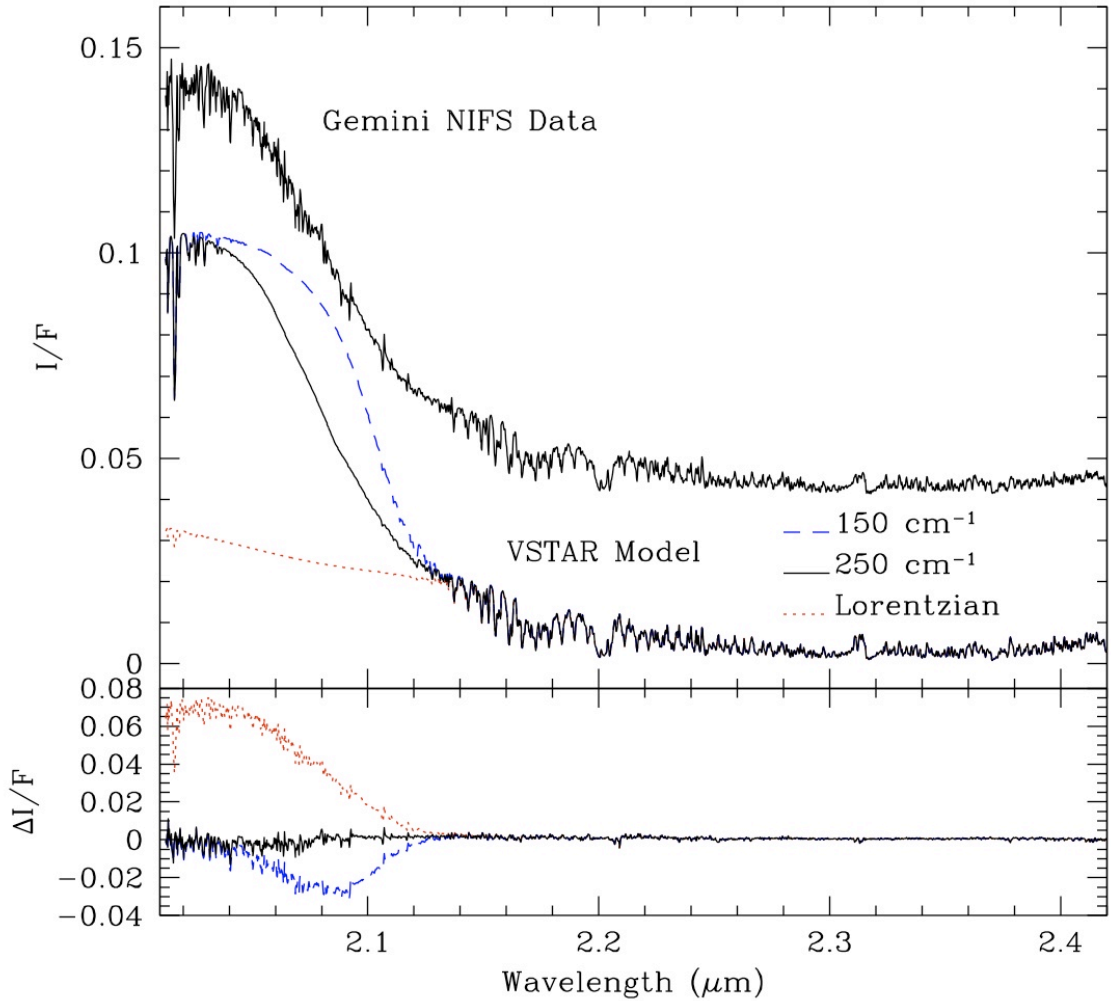

*Figure 8 – Observed NIFS spectrum of Titan in the 2.0 μm window (shifted up by 0.04) compared with model spectra calculated with different far-wing line shapes. The dashed line (blue) uses the line shape model of equation 2 that fits the 1.55 μm window. The dotted line (red) uses a Lorentzian profile continued to large distances from the core. The solid line is the line shape given in equation 3, which is Lorentzian to 250 $cm^{-1}$, and then becomes sub-Lorentzian.*

## 6.3 CO and CH$_3$D in the 1.55 μm Window

In addition to methane absorption the 1.55 μm window includes significant absorption features due to CH$_3$D, and absorption due to the 3-0 band of CO in the 1.575 μm region. Figure 9 shows the residuals of the data when fitted with a model that includes CH$_4$ lines, but no CH$_3$D or CO lines. Absorptions of CH$_3$D are apparent over the 1.54 to 1.56 μm region where the 3ν$_2$ band is important, and also at 1.588 μm where we see the ν$_1$+ν$_2$+ν$_6$ band (Ulenikov et al., 2010).

Our line list, in this region, is based on the data of Wang et al. (2010b) and their analysis has taken care to identify lines of CH$_3$D. Hence we can use the lines in this region to determine the D/H ratio of methane in Titan's atmosphere. We calculated a set of models in which the D/H ratio was varied from 0.4 to 1.5 times the terrestrial value in steps of 0.1, and used linear interpolation between the spectra for intermediate values. We compared the models with the data in several wavelength regions that contain strong CH$_3$D lines. The best fitting D/H ratio was determined for each of these wavelength regions, and each of the four individual observations that make up the H band data cube, yielding 12 independent measurements of the D/H ratio. From these measurements we determine an average D/H of 1.147 ± 0.13 times terrestrial, or a D/H of 1.77 ± 0.20 × 10$^{-4}$. The uncertainty of this determination is dominated by the differences between measurements in different wavelength regions, which indicate remaining uncertainties in the CH$_3$D or CH$_4$ line data or other aspects of the model. The statistical differences between observations were much smaller. The third panel of Fig. 9 shows the residuals after CH$_3$D has been included in the model.

In a similar way we have determined the CO mixing ratio by calculating a set of models in which CO is varied from 30 to 84 ppmv in 6 ppmv steps. The best fitting CO mixing ratio was determined by fitting wavelength regions including CO lines from 1.57 to 1.584 μm after the D/H value had already been determined (as there are CH$_3$D lines in this region). The best fitting CO mixing ratio is 50 ppmv with a statistical error of 4 ppmv. However, the CO determination (unlike the D/H one) is dependent on our adopted CH$_4$ mixing ratio (of 4.92 % in the lower atmosphere) since we have adjusted the path length in our model to give the correct CH$_4$ absorption with this CH$_4$ mixing ratio. We have therefore increased the uncertainty of our CO mixing ratio to allow for a 1% uncertainty in the CH$_4$ mixing ratio, giving a value of 50 ± 11 ppmv. The lower two panels of Fig. 9 shows the residuals after both CH$_3$D and CO have been included in the model.

There are two apparent absorption features that remain in the residuals at wavelengths of 1.5529 and 1.5606 μm. These appear to be real features as they are present in all our individual spectra, but we cannot identify them with known absorptions of any likely constituent of the Titan atmosphere, nor with telluric or solar lines that might have been incompletely corrected. The most likely explanation is that they are methane lines that are missing from, or have incorrect data, in the line list. Ignoring these two features the RMS residuals of the final model fit to I/F are 0.0015 and 0.0023 in the CH$_3$D and CO regions respectively. This means our model is fitting the spectrum to 2-2.5% in this region, which is an indication of the quality of the line list.

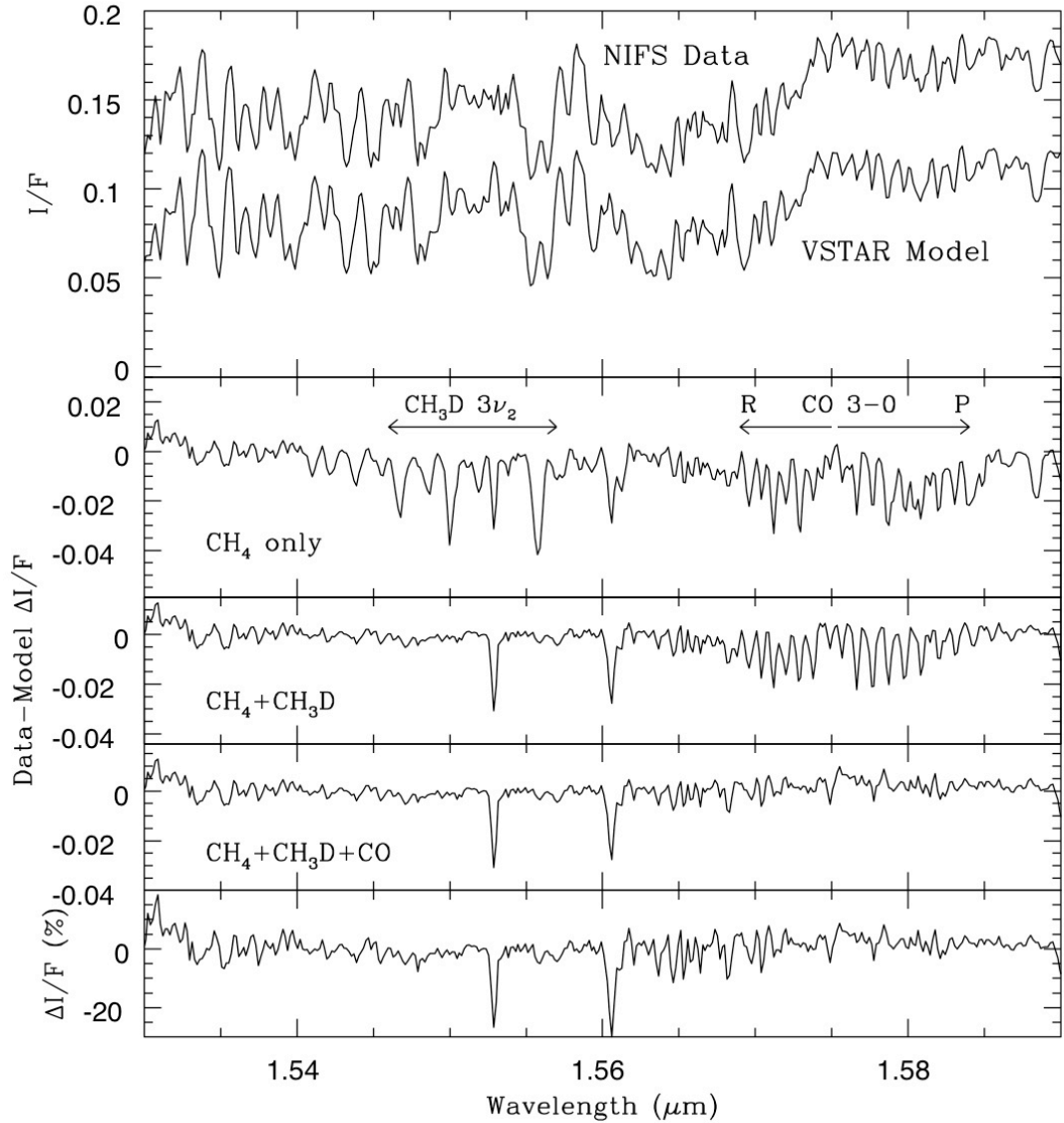

*Figure 9 — NIFS spectrum compared with model in the 1.53 to 1.59 μm wavelength region. The upper panel shows the observed spectrum (shifted up by 0.06) compared with the modeled spectrum. The second panel shows the residuals of a model including only $CH_4$ lines. Absorption features due to the R and P branches of the CO 3-0 band, as well as absorptions due to $CH_3D$ can be seen. The third panel adds $CH_3D$ lines to the model based on our best fitting D/H ratio, and the fourth panel adds CO lines with a mixing ratio of 50 ppmv. The lower panel shows the residuals of the final model as a percentage of the modelled spectrum.*

## 6.4 $CH_3D$ in the 2 μm Window

Comparison of the NIFS data with the model in the 2.01 to 2.06 μm region shows a number of absorption features that do not appear to be due to $CH_4$ lines in our line list. Fig. 10 shows the observed spectrum in this region compared with our modeled spectrum. A number of lines at the left hand end of this region are consistent with the model. These lines are due to the $4\nu_4$ band of $CH_4$ and lines in this part of our line list come from the STDS model, which in this region has been fitted to experimental data for the $4\nu_4$ band by Robert et al. (2001). While the line list includes $CH_4$ lines throughout this region, at longer wavelengths these lines rapidly become too weak to detect, because these are high-J transitions that have very low intensities at the low temperatures of the Titan atmosphere.

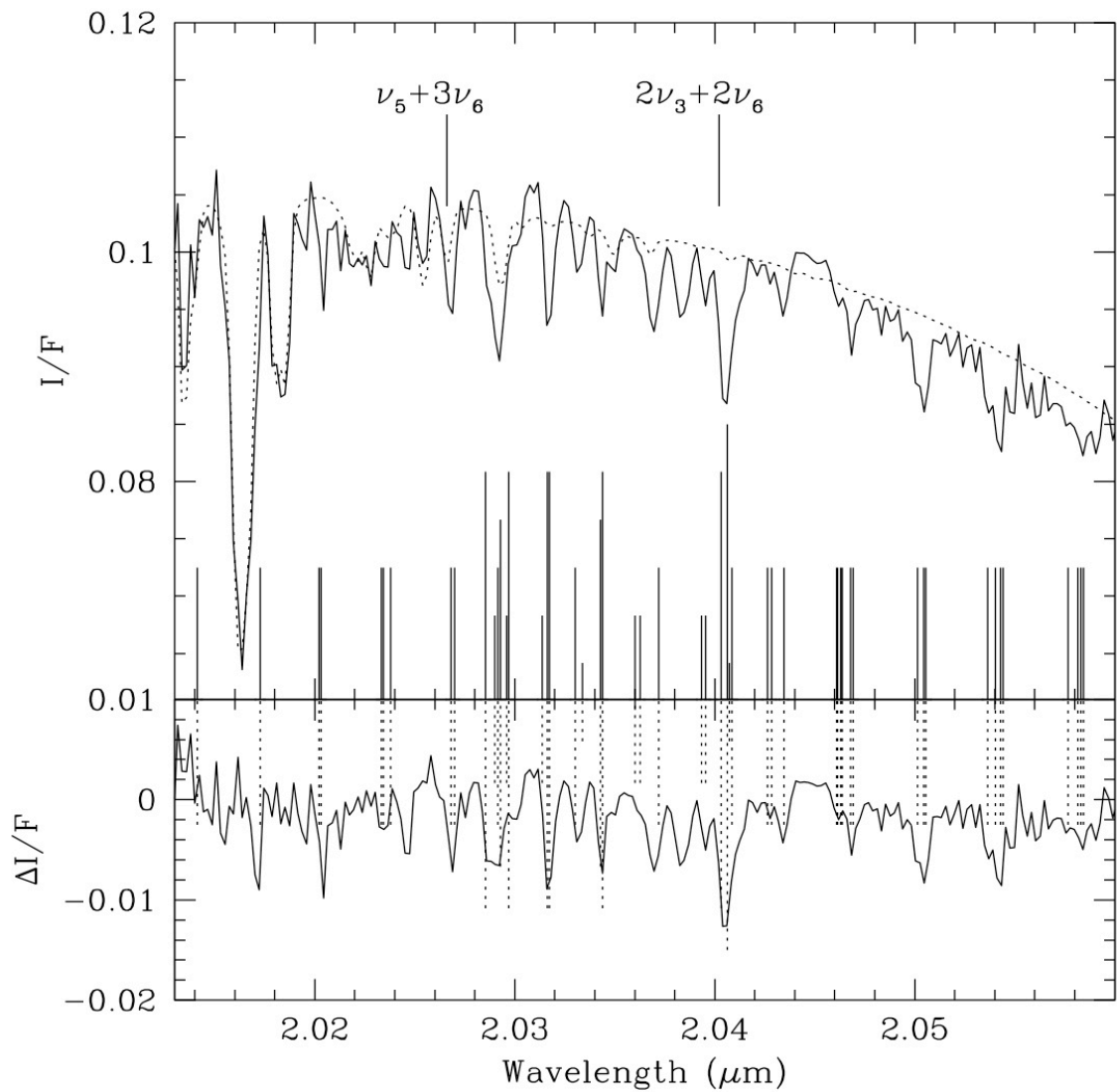

*Figure 10 — Observed spectrum (solid line) and VSTAR model including $CH_4$ absorption (dotted line) in the 2 μm window. The residuals (Data-Model) are shown in the lower panel. There are many absorption features in this region that are not fitted with $CH_4$ lines in our model. The "stick spectrum" shows the positions of $CH_3D$ lines in this region from Ulenikov et al. (2010). The band centres of the $CH_3D$ $\nu_5+3\nu_6$ and $2\nu_3+2\nu_6$ bands are also indicated.*

The observed spectrum, however, shows a number of distinct absorption features in this region the strongest being at wavelengths of 2.029 and 2.04 μm. Although no $CH_3D$ lines in this region are included in HITRAN, a recent analysis (Ulenikov et al., 2010) has reported the experimental detection in Fourier Transform spectra at 80 K, of the $\nu_5+3\nu_6$ and $2\nu_3+2\nu_6$ bands of $CH_3D$ with band centres at 2.027 and 2.040 µm. Ulenikov et al., 2010 have published a list giving line positions (but not line intensities). Approximate relative line intensities for part of this range can be estimated from their figure 6. These data have been used to plot the "stick spectrum" shown at the bottom of Fig. 10. It can be seen that virtually all the lines in our Titan spectrum in this region correspond to the positions of lines of these two $CH_3D$ bands. Since we do not have line intensities for these lines we cannot currently include them in our model.

# 7 Discussion

The modeled spectra are in good agreement with the observations at all wavelengths longer than 1.3 µm. In particular the new line list provides a spectrum in the 1.55 µm transparency window that fits the observed spectrum to an accuracy of 2-2.5%, whereas this region could not be modeled usefully using HITRAN 2008. The spectra are a significant improvement on those obtained with HITRAN 2008 at all wavelengths from 1.3 to 2.1 µm, fitting the absorption band regions to accuracies of about 8%. While the line list is a substantial improvement on what has been available previously, there are still regions of the spectrum (in particular at 1.49 to 1.53 µm and 1.59 to 1.62 µm) that lack good low-temperature measurements and where further observations would be valuable. It would also be useful to extend the line list to shorter wavelengths so that the 1.25 µm window could also be modeled.

To correctly fit both the I/F levels and the depth of the methane absorption in the windows and deep absorption bands, we needed to modify the aerosol optical depths from the values derived by Tomasko et al. (2008) from Huygens DISR data. In the highest altitude range (above 80 km) we used a scaling factor of 0.8. Given that our spectra lack absolute calibration, and that we used a simplified version of the scattering phase function this should probably not be considered a significant difference. However the scaling factors of 0.5 for the 30-80 km range and 0.3 for the below 30 km region suggest a significant change in the aerosols at the time of our observations compared with that at the time of the Huygens descent just over a year earlier. What we are sensitive to is the amount of back scattered light from the aerosols which provides diluting radiation that reduces the depth of the absorption features compared with what would be seen if all the light in the window regions was reflected from the surface. Thus although we have adjusted this in our model by changing the aerosol optical depth, a similar effect could be obtained by changing other aerosol properties such as the asymmetry parameter or single scattering albedo.

The ability to model these regions of the spectrum at high resolution using line-by-line methods will be valuable for a number of studies. A good model for the methane lines improves our ability to detect and measure weaker spectral features and this is shown in our data by the ease of detection of $CH_3D$ and CO lines in the 1.55 µm region after the $CH_4$ lines are modeled.

We have used our spectra to determine the CO mixing ratio, and the $CH_3D/CH_4$ ratio. The CO mixing ratio we measure is 50 ± 11 ppm. CO is thought to have a long chemical lifetime in Titan's atmosphere, and so would be expected to be uniformly mixed in the atmosphere (Lellouch et al., 2003). Our value is in good agreement with determinations of the stratospheric CO mixing ratio of 51 ± 4 ppm (Gurwell et al., 2004) and 47 ± 8 ppm (de Kok et al., 2007), but agrees less well with determinations of 27 ± 5 ppm in the stratosphere (Hidayat et al. 1998) and 32 ± 10 ppm in the troposphere (Lellouch et al., 2003).

Our D/H ratio determination of $(1.77 ± 0.20) \times 10^{-4}$ is towards the high end of the range of previous measurements (summarized in Abbas et al., 2010) that range from $0.775 \times 10^{-4}$ to $2.3 \times 10^{-4}$. It is in good agreement with the recent determination of $(1.58 ± 0.16) \times 10^{-4}$ from Cassini CIRS (Abbas et al., 2010) and the in-situ measurement from Huygens of $(2.3 ± 0.5) \times 10^{-4}$ (Niemann et al., 2005). The measurement confirms previous findings that the Titan D/H ratio is substantially enriched compared with the estimated protosolar value of $2 \times 10^{-5}$ (Geiss & Reeves, 1981).

It is important to note that we have been able to make good measurements of CO and D/H with much lower resolution spectra than have generally been used for this purpose in the past, thanks to the ability to accurately model the $CH_4$ lines using our new line list. It should be possible to obtain even better measurements using higher resolution spectra of Titan.

We have also shown that spectral features in the 2 μm window are due to the $\nu_5+3\nu_6$ and $2\nu_3+2\nu_6$ bands of $CH_3D$. These lines are easily seen in our spectra with a resolving power of ~5000, but would be strong features at even higher spectral resolution. Because these features occur in a region of the spectrum that is largely clear of $CH_4$ lines (but there are $CH_4$ lines at slightly shorter wavelengths that can be used for comparison) it should be possible to measure the D/H ratio of Titan and the giant planets with high accuracy using these lines. However, we need good laboratory data for the intensities and lower state energies of these $CH_3D$ lines to be able to add them to our line list.

We note that Negrao et al. (2007) derived a surface albedo that decreased sharply from about 0.14 at 2.03 μm to 0.02 at 2.11 μm. Such a sharp decrease in albedo is hard to explain based on the known properties of ices that might be present on the Titan surface. We suggest that the apparent decrease was the result of the use of an inappropriate far-wing line shape model for methane. Our results show that good agreement with the observations can be obtained with an albedo that has no wavelength dependence if a suitable line shape model is used. Our model uses a surface albedo that is constant at 0.096 throughout the 2 μm window. Without good constraints on the methane far-wing line shape from independent data, it is not easy to distinguish these two effects, and it is therefore difficult to constrain the wavelength dependence of the surface albedo.

## 8   Conclusions

We have constructed a new spectral line list for methane based on new data including recent low-temperature cavity ring down spectroscopy data in the 1.3 to 1.8 μm spectral region. The line list goes substantially deeper than the HITRAN 2008 list in these regions and includes many more lines with measured lower state energies, which are needed to predict line intensities at Titan temperatures.

We have used the new line list in conjunction with a line-by-line, multiple scattering radiative transfer model to predict the Titan spectrum, and compared this with observed spectra of Titan obtained with NIFS on the Gemini North 8m telescope. The atmospheric structure and composition, aerosol properties and surface albedo were constrained with the help of in-situ measurements from the Huygens probe.

With the new line list we are able to obtain spectra that provide a good match to the observations over the wavelength range from 1.3 to 2.43 μm. In particular we are now able to obtain good model spectra for the window region at 1.55 μm, which could not be modeled well with previous line lists.

The far wing line shape is important in determining the spectral shape of the window regions between the strong methane bands. The 1.55 μm window is fitted with a line shape that is Lorentzian out to 150 cm$^{-1}$, and sub-Lorentzian at larger distances. However, the same line shape does not provide a good fit to the 2 μm window and a line shape with a wider Lorentzian region is needed here.

After fitting the $CH_4$ spectrum, absorptions due to other species are easily visible, and in the 1.55 mm region we can easily detect absorptions due to CO and $CH_3D$, and we use these to measure the CO mixing ratio and D/H ratio. In the 2.0 μm window we show that there are absorption lines that can be identified with the $v_5+3v_6$ and $2v_3+2v_6$ bands of $CH_3D$. This wavelength region will also be valuable for measuring D/H ratios in Titan and the giant planets when good laboratory data on these lines becomes available.

The new line list makes it possible to investigate the spectra of Titan and the solar system giant planets at high spectral resolution, and better understand their atmospheric structure and composition, by enabling a search for spectral features of other trace constituents.

The methane line list, in HITRAN 2004 format, is available for download from http://www.phys.unsw.edu.au/~jbailey/ch4.html.

## Acknowledgments
We thank Alain Campargue for drawing our attention to the new low-temperature methane line measurements and Vincent Boudon, Jean Paul Champion and Christian Wenger for their assistance with running the STDS package. Based on observations obtained at the Gemini Observatory, which is operated by the Association of Universities for Research in Astronomy, Inc., under a cooperative agreement with the NSF on behalf of the Gemini partnership: the National Science Foundation (United States), the Science and Technology Facilities Council (United Kingdom), the National Research Council (Canada), CONICYT (Chile), the Australian Research Council (Australia), Ministério da Ciência e Tecnologia (Brazil) and Ministerio de Ciencia, Tecnología e Innovación Productiva (Argentina). The observations were obtained as part of the system verification of the NIFS instrument under program GN-2006A-SV-128. We thank Tracy Beck for coordinating the observations at Gemini and for assistance with the data reduction scripts. Tom Geballe set up the Phase II observing sequences for the observations. We thank Dr L. Sromovsky and an anonymous reviewer for suggestions that substantially improved this paper.